# Conductance oscillations in tunnel-coupled quantum dots in the quantum Hall regime

C. Livermore*, D.S. Duncan, and R.M. Westervelt

*Division of Engineering and Applied Sciences and Department of Physics, Harvard University, Cambridge MA 02138*

K.D. Maranowski and A.C. Gossard

*Materials Department, University of California, Santa Barbara CA 93106*

Abstract

We present measurements of transport through two tunnel-coupled quantum dots of different sizes connected in series in a strong, variable, perpendicular magnetic field. Double dot conductance was measured both as a function of magnetic field, which was varied across the filling factor $\nu = 4$ quantum Hall plateau, and as a function of charge induced evenly on the two dots. The conductance peaks undergo position shifts and height modulations as the magnetic field is varied. These shifts and modulations form a pattern that repeats over large ranges of magnetic field and with the addition of double dot charge. The robust pattern repetition is consistent with a frequency locking effect.

Transport through single quantum dots in a strong, perpendicular magnetic field depends sensitively on the magnetic field value [1-9]. For example, as magnetic field increases electrons may be redistributed from higher to lower Landau levels [1, 2, 4, 7, 8]. The frequency of such redistributions is related to the change in field required to thread an additional flux quantum through the dot area. Charge redistributions occur when the lowest energy configuration of the system changes, and they are clearly reflected in modulations of conductance peak height and shifts in conductance peak position.

In this experiment we measured how conductance through two tunnel-coupled quantum dots in series changes with small variations of a strong perpendicular applied magnetic field and with the addition of charge. This has not been examined before, and these experiments raise a number of interesting questions. The device we studied consists of two quantum dots of different lithographic sizes connected in series via an adjustable quantum point contact. A strong perpendicular magnetic field places the electron gas in the quantum Hall regime; for the measurements described here, the electron gas is on the quantum Hall plateau corresponding to a Landau level filling factor $\nu = 4$. By measuring conductance through the double dot as a function of the induced dot charges while sweeping the magnetic field across the quantum Hall plateau, we monitor the electron readjustments that trace out changes in the double dot ground state.

Figure 1 is a scanning electron micrograph image of our device structure, the asymmetric double dot. The device is composed of metal gates on the surface of a GaAs/AlGaAs heterostructure wafer that contains a two-dimensional electron gas (2DEG). The device wiring is indicated on the surface gates. The 2DEG mobility $\mu \cong 450{,}000\ cm^2V^{-1}s^{-1}$, the sheet density $n_s \cong 3.4 \times 10^{11}\ cm^{-2}$, and the depth of the gas beneath the wafer surface is 57 nm. When all ten gates are energized, two quantum dots are defined in series. Varying the side gate voltages $V_{g1}$ and $V_{g2}$ together induces charge simultaneously on the two dots; adjusting the center point contact gate voltage $V_{q2}$ controls the interdot tunnel conductance $G_{int}$. The lithographic sizes of the dots, 500 nm x 600 nm and 500 nm x 800 nm, combine with about 75 nm of depletion width beyond the gates to produce dot areas approximately 0.16 $\mu m^2$ and 0.23 $\mu m^2$ respectively. The area asymmetry is crucial for these experiments because these quantum dots have different characteristic frequencies for the addition of flux quanta to the dot area, as well as different higher harmonics.

We cool the device to base temperature (30 mK) in a dilution refrigerator and energize the magnet to produce a perpendicular field B ≅ 3.2 T, corresponding to a filling factor $\nu = 4$. A small ac bias voltage (5 $\mu V_{rms}$) is applied across the device and the resulting current is measured using a current preamplifier with lockin detection. The capacitance $C_{g1}$ ($C_{g2}$) coupling the gate labeled $V_{g1}$ ($V_{g2}$) to dot 1 (dot 2) is found from measurements of dot conductance vs. $V_{g1}$ ($V_{g2}$) to be $C_{g1} \cong 45 \pm 4$ aF ($C_{g2} \cong 43 \pm 4$ aF).

The characteristic magnetic field frequencies for the addition of flux quanta to each dot were calibrated by energizing a subset of the surface gates to form a single quantum dot and measuring its conductance as a function of magnetic field and induced charge. Figure 2(a) plots the conductance of dot 1, measured alone, vs. gate voltage $V_{g1}$ on the vertical axis and the perpendicular magnetic field B on the horizontal axis. Conductance is plotted in inverted logarithmic grayscale, with dark indicating high conductance and light indicating low conductance. The magnetic field range is chosen to remain on the $\nu = 4$ plateau. Figure 2(b) is a similar plot of dot 2 conductance vs. side gate voltage $V_{g2}$ and magnetic field. The data were obtained by repeatedly measuring dot conductance as a function of side gate voltage while continuously varying the magnetic field. The conductance peak positions shift in a periodic zigzag pattern that correlates with peak height modulations. The peaks move gradually as the magnetic field shifts the energy of the dot; the sudden return to the initial peak position signals that electrons have been redistributed among the Landau levels to minimize the energy. Peak modulations have been observed previously for single dots [1, 2, 4, 7, 8]. For the dots described here, the shifts in peak position are periodic in magnetic field.

The periods of the zigzag patterns for the two dots are different, corresponding to the distinct dot areas. The period of the modulations on the $\nu = 4$ quantum Hall plateau is 6.6 mT for dot 1 and 8.75 mT for dot 2. We can calculate the period $\Delta B$ corresponding to threading one flux quantum through the known dot areas from the relation $\Delta B A_{dot} = h/e$. With a 75 nm depletion length as assumed above, this yields periods $\Delta B = 18.2$ mT for dot 1 and $\Delta B = 26.2$ mT for dot 2, which are 2.8 and 3.0 times the observed periods for dots 1 and 2 respectively. The short periods are observed because the number of electrons in each of the lower Landau levels increases every time a flux quantum is threaded through its area.

We then measured the evolution of double dot conductance with magnetic field to investigate how charge redistributions minimize the double dot energy in a magnetic field. All ten gates were energized to place the double dot in the Coulomb blockade regime. Both outer point contacts were set in the tunneling regime, with conductance $G_{q1}, G_{q3} << e^2/h$. The interdot tunnel conductance was calibrated independently and set to $0 \le G_{int} \le e^2/h$ [10]. We repeatedly measured the conductance through the entire double dot as a function of the side gate voltages $V_{g1} = V_{g2}$, which induce charge simultaneously on both dots, while continuously varying the magnetic field. Varying $V_{g1} = V_{g2}$ is equivalent to taking a diagonal slice through the double dot charging diagrams measured in previous experiments [11-18]. We repeated this measurement for a series of interdot conductances ranging from $G_{int} \cong 0$ to $G_{int} \cong e^2/h$.

Figures 3(a)-(d) plot double dot conductance in inverted logarithmic grayscale as a function of the side gate voltages $V_{g1} = V_{g2}$ on the vertical axis and magnetic field on the horizontal axis. In panels (a) through (c), the gate voltages have been adjusted such that $V_{g1} = V_{g2}$ crosses the points in the charging diagram where the Coulomb blockade is simultaneously lifted for both dots; in (d), they are set such that $V_{g1} = V_{g2}$ crosses the points where the Coulomb blockade is lifted for only one dot at a time. Figure 3(a) corresponds to $G_{int} \cong 0.20\ e^2/h$. A vertical slice through the data reveals a series of narrowly split peaks, as is expected for a double dot with low interdot conductance. As the magnetic field is varied, these peaks evolve in both position and height. The structure is qualitatively similar to that for the single quantum dots; peak amplitude is strongly modulated, and we observe a zigzag pattern in the peak positions. Unlike the single dot case, the zigzags are no longer evenly spaced, and there are blank spots where the conductance is strongly suppressed. The peak modulations/shifts and blank spots form a repeated pattern that is marked by the series of letters A - F in Fig. 3(a). The pattern is more clearly visible for higher values of interdot conductance and will be discussed further in that context.

Figures 3(b) and 3(c) plot double dot conductance vs. $V_{g1} = V_{g2}$ and magnetic field for interdot conductances of $G_{int} \cong 0.85\ e^2/h$ and $G_{int} \cong 1.03\ e^2/h$ respectively. As a given peak evolves with magnetic field, the shifts and height modulations form a pattern in which a given sequence repeats with magnetic field. The repeating sequence consists of six distinct peak position shifts, which are marked in Figs. 3(b) and (c) by the series of letters A, B, C, D, E, and F. The magnetic field scales of the shifts are consistent with the pattern marked in Fig. 3(a) as well, though at low interdot conductance the pattern is more visible in peak height modulations than in peak shifts. The length of the repeated segment is measured at 25.6 mT, which is $3.9 \pm 0.1$ times the observed period for dot 1 and $3.0 \pm 0.2$ times the measured period for dot 2 [19]. The periods are nearly constant with changing interdot tunnel conductance; small variations are expected because the dot areas change as the center point contact voltage is adjusted. In Fig. 3(a)-(c), we see about four repetitions of the pattern as we track a given peak in magnetic field; measurements over a broad field range confirm that the pattern repeats across a given quantum Hall plateau.

The pattern in Fig. 3(a)-(c) also appears robust to the addition of double dot charge. The two component peaks that comprise a single split peak undergo the same pattern of peak modulations, as do subsequent pairs of split peaks. However, the corresponding peak shifts do not occur at exactly the same magnetic field. Figure 3(a)-(c) shows that the pattern repeats at higher magnetic field and at less negative gate voltage (or at lower magnetic field and more negative gate voltage). The pattern is shifted both within a given peak pair and between neighboring pairs, for an overall visual effect of sloping bands of corresponding peaks. We interpret correspondence of peaks as meaning that the double dot has similar structure at those values of field and gate voltage. Increasing the field decreases the spatial size of the flux quanta and packs the electrons more tightly. Adding additional charge compensates by increasing the radius.

The consistent pattern of Fig. 3(a)-(c) indicates that when the Coulomb blockade is lifted for both dots, the energy of the system is minimized by a single, specific sequence of adjustments in the electron distribution that repeats each time four additional flux quanta are added to the large dot. The energy minimization sequence appears independent of magnetic field (for a given filling factor) and repeats periodically at a single frequency characteristic of the entire double dot. This is in direct contrast to the quasiperiodic readjustment events that one might expect for independent processes on two separate quantum dots. Experiments on single dots have found that transport features may correlate over ranges of magnetic field or gate voltage [9, 20]. It is noteworthy that the more complicated double dot system also locks into a long range pattern.

Figure 3(d) plots double dot conductance vs. $V_{g1} = V_{g2}$ and magnetic field at an interdot conductance $G_{int} \cong 0.66\ e^2/h$. In these data, $V_{g1} = V_{g2}$ crosses the charging diagram at the points where the Coulomb blockade is lifted for only one dot at a time. This is evidenced both by the reduced Coulomb blockade peak height and by the peak splitting in the $V_{g1} = V_{g2}$ direction, which is larger in (d) than in (b) despite (d)'s smaller interdot tunnel conductance. In Fig. 3(d) the two peaks that comprise a split peak undergo modulations at different frequencies, in contrast with the repeated pattern seen in Fig. 3(a)-(c). The observed frequencies are the same as those determined from the single dot measurements of Fig. 2. The two independent frequencies indicate that when the Coulomb blockade is lifted for just one of the dots, the energy of the system is minimized by readjusting the electrons on that dot alone rather than on the entire double dot. The electrons are sufficiently shared between the two dots to permit current flow; however, the time scale for charge readjustments involving the blockaded dot appear to be longer than the time scale for charge sharing between dots.

Figure 4 illustrates a likely source of the pattern of double dot conductance modulations observed in Fig. 3(a)-(c). Recall that the conductance peaks for individual dots shift and undergo peak height modulations as the magnetic field varies. This behavior is shown schematically by the sets of solid lines labeled dot 2 and dot 1, which represent conductance peaks; the peaks shift at different frequencies for the two different sized dots. If we superpose these two structures, postulating that the double dot conductance is modulated each time a single dot undergoes a conductance modulation, we obtain the sequence of modulations shown in the third line. This sequence forms a repeated pattern only to the extent that the phase relation is fixed with exactly four modulations in the large dot for every three in the small dot, as shown. Otherwise, the pattern slips steadily away. The sequence obtained in this way is similar to the observed double dot conductance pattern, and segments are marked with the letters A - F.

The repeated pattern of double dot conductance modulations suggests a frequency locking effect, in which interdot interactions drive periodic electron redistributions at a frequency that is close to a commensuration frequency for the two dots. In the simplest picture of non-interacting dots, the conductance modulations would be quasi-periodic and would form a repeated pattern only to the extent that the ratio of the two periods is rational. The measured ratio of the periods of the two dots is $0.77 \pm 0.06$, which is approximately a rational ratio of three to four. Although it would be possible to observe this repeated pattern if the frequency ratio were perfectly rational, it could be difficult to maintain because the periods do vary with changing gate voltages. A more likely explanation of the repeated pattern is that the double dot conductance modulations are not just the sum of the single dot modulations, but rather lock onto the commensuration frequency corresponding to the almost rational frequency ratio of the individual dots.

We thank P. Brouwer and B.I. Halperin for helpful discussions and comments on the manuscript and R.G. Beck, D. Bozovic, L. Chen, D. Davidovic, M. Drndic, S. Pohlen, and M.A. Topinka for experimental assistance. Supported at Harvard by NSF grant DMR-95-01438, ONR grant N00014-95-1-0866, and the MRSEC program of the NSF under award DMR-94-00396; supported at UCSB by grant AFOSR F49620-94-1-0158 and by QUEST; and D.S.D. acknowledges support by the DOD NDSEG Fellowship program.

obtained from this measurement by adding a calibrated lateral shift to account for the cross capacitance between the outer point contact gates and the center point contact.

[19] The observed periods for the individual dots shift when the full double dot is energized because of cross capacitance from the additional gates. The resulting periods are $6.55 \pm 0.15$ mT for dot 1 and $8.55 \pm 0.5$ mT for dot 2.

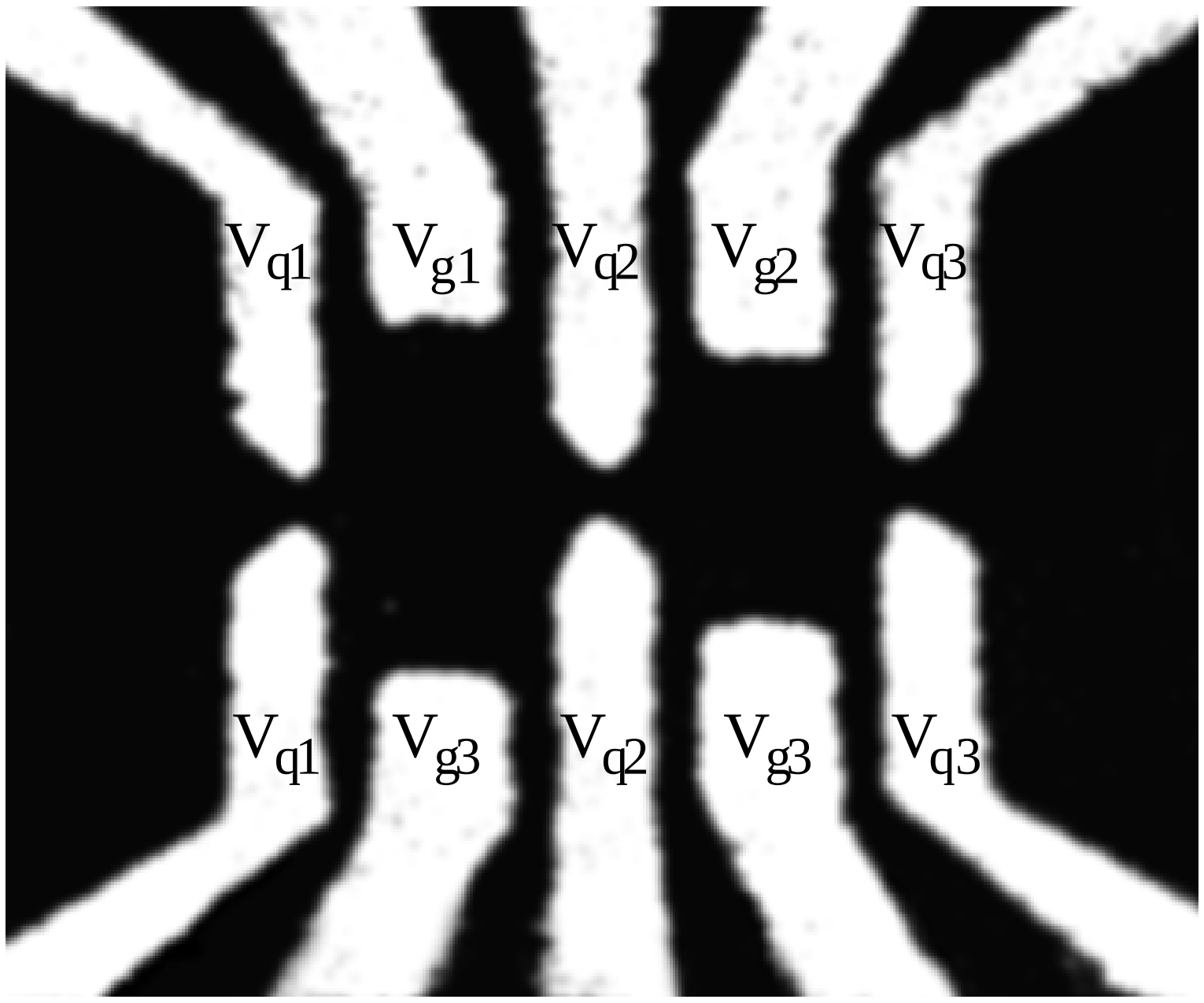

Figure 1

C. Livermore *et al.*

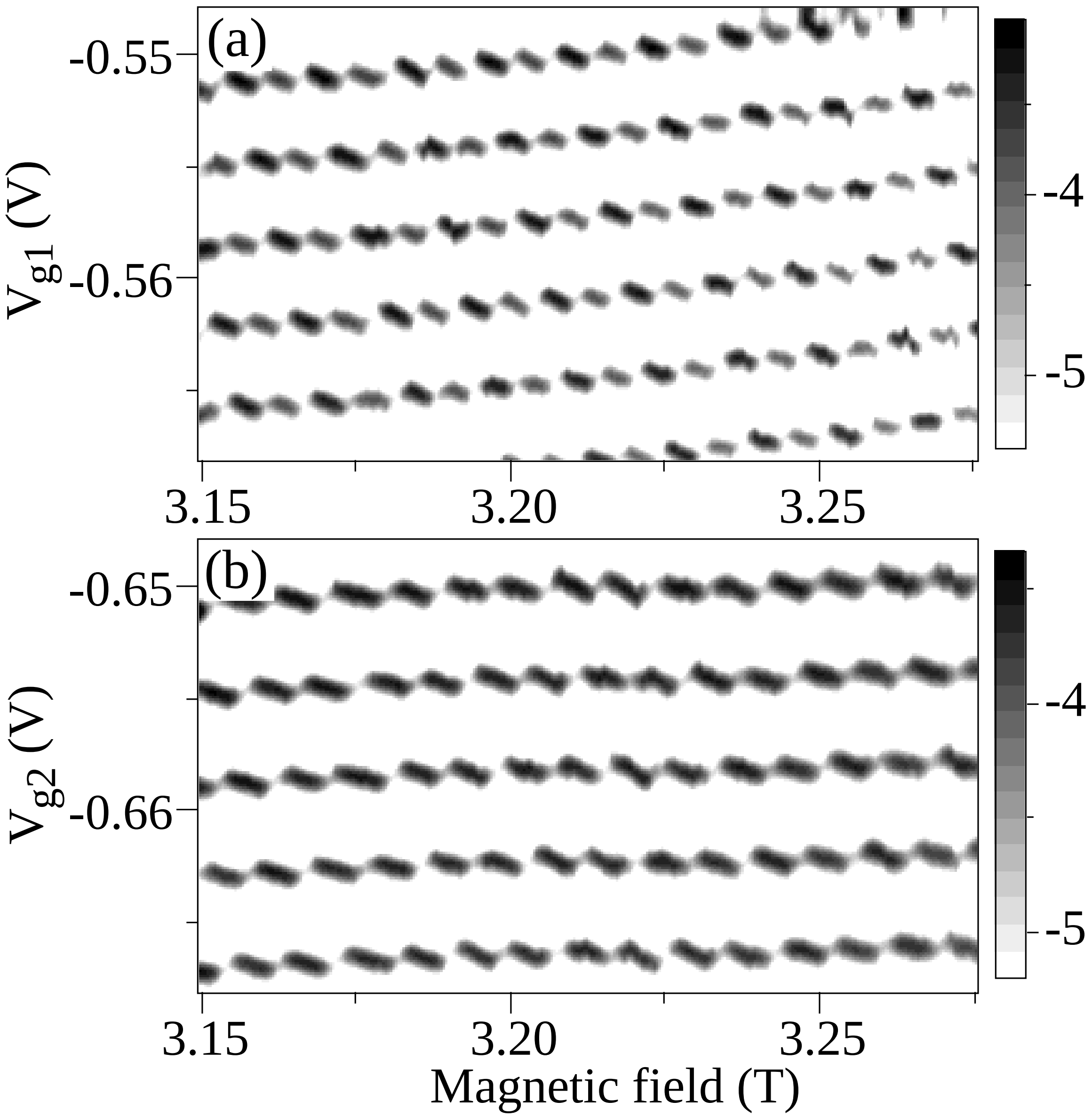

Figure 2

C. Livermore *et al.*

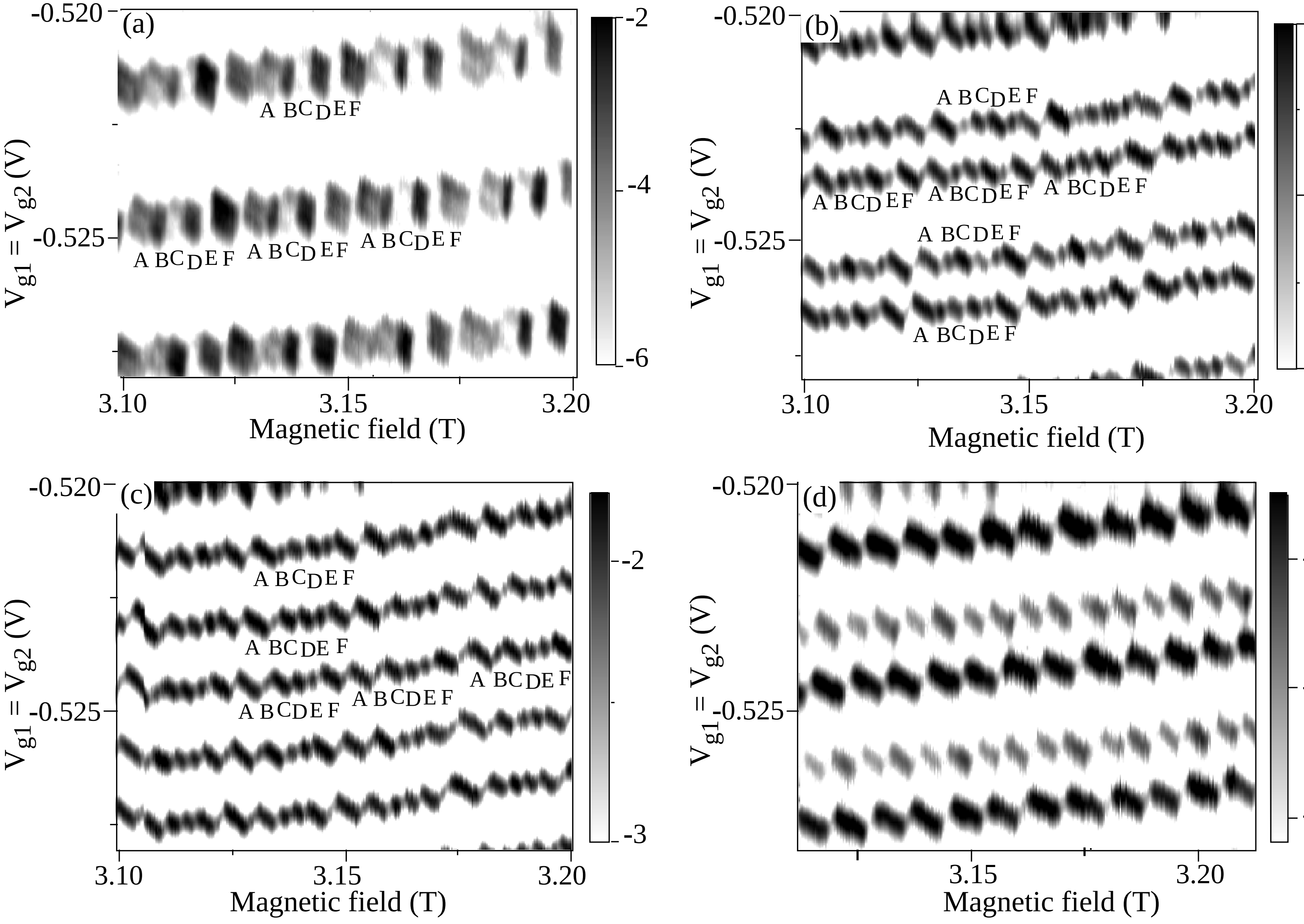

Figure 3

C. Livermore *et al.*

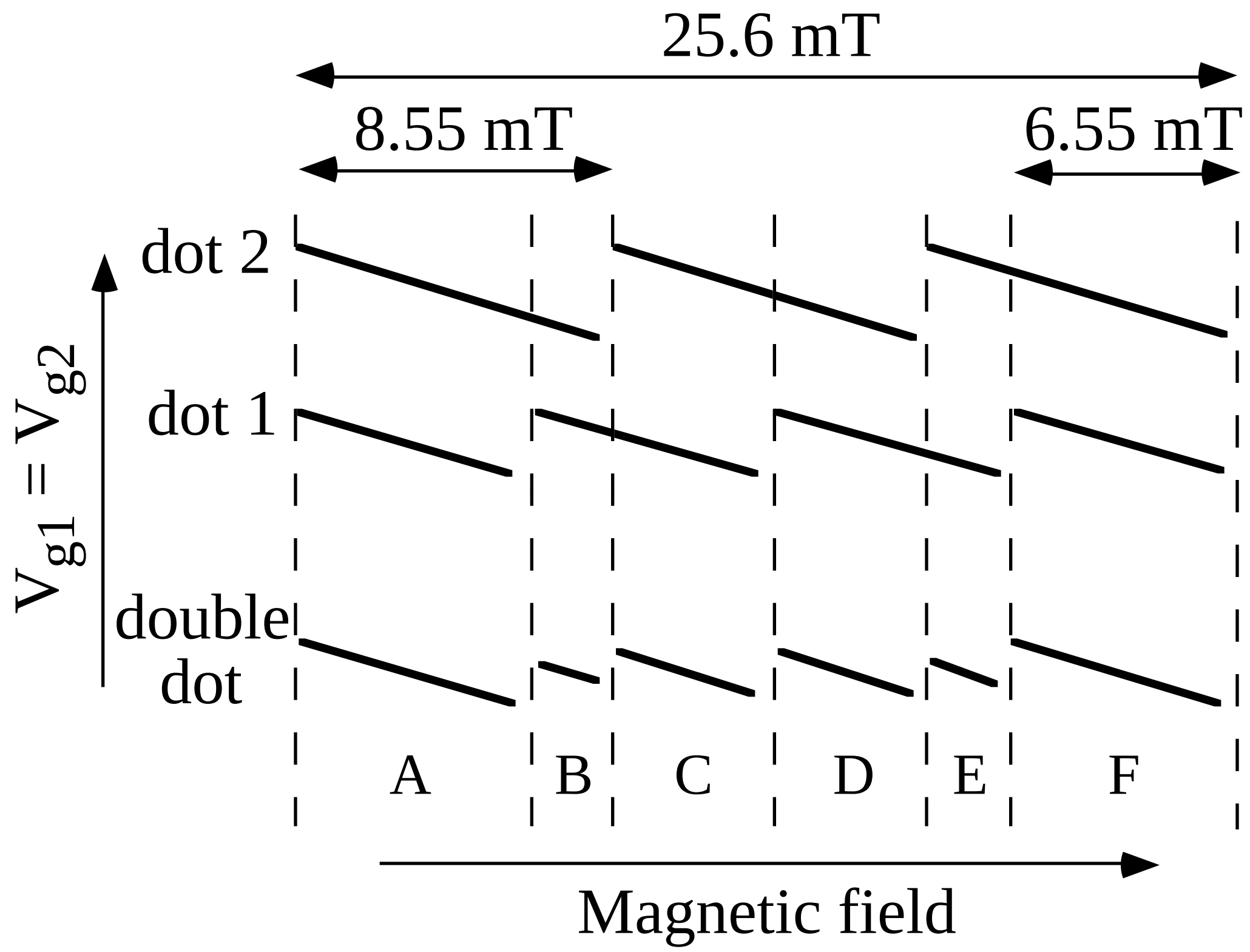

Figure 4

C. Livermore *et al.*

## Figure captions

FIG. 1. Scanning electron micrograph of an asymmetric double quantum dot device. Light regions are the metal surface gates; dark regions are the wafer surface. The gate wiring is indicated.

FIG. 2. (a) Inverted grayscale image of the logarithm of conductance through dot 1, measured alone in units of $e^2/h$, plotted vs. side gate voltage $V_{g1}$ on the vertical axis and perpendicular magnetic field B on the horizontal axis. Conductance peak heights and positions are modulated as electrons are redistributed to minimize the energy of the dot; the period of these modulations is related to dot area. (b) Similar image of logarithm of conductance through dot 2 plotted vs. $V_{g2}$ on the vertical axis and B on the horizontal axis. The period is different, corresponding to the different dot area.

FIG. 3. (a)-(d) Inverted grayscale images of the logarithm of double dot conductance (measured in units of $e^2/h$) plotted vs. side gate voltages $V_{g1} = V_{g2}$ on the vertical axis and perpendicular magnetic field B on the horizontal axis for interdot tunnel conductance (a) $G_{int} \cong 0.20\ e^2/h$, (b) $G_{int} \cong 0.85\ e^2/h$, (c) $G_{int} \cong 1.03\ e^2/h$, and (d) $G_{int} \cong 0.66\ e^2/h$. Color bars summarize the grayscale for each panel. In (a) – (c) $V_{g1} = V_{g2}$ crossed the charging diagram at points where the Coulomb blockade was lifted for both dots. The repeated pattern of conductance modulations in (a) – (c) is marked by the letters A - F. In (d) $V_{g1} = V_{g2}$ crossed the charging diagram at points where the Coulomb blockade was lifted for only one dot at a time.

FIG. 4. Schematic diagram of a possible mechanism for the observed double dot conductance modulations. The top two lines represent conductance modulations of individual dots; bottom line represents a double dot conductance modulation occuring

whenever either dot undergoes a conductance modulation. The resulting pattern A - F is marked.